# Cosmic Ray Electrons and Protons ~1 MeV - A 40 Year Study of Their Intensities from the Earth to the Heliopause and Beyond Into Local Interstellar Space by the CRS Experiment on Voyager 1

W.R. Webber[1], N. Lal[2] and B. Heikkila[2]

1. New Mexico State University, Astronomy Department, Las Cruces, NM 88003, USA
2. NASA/Goddard Space Flight Center, Greenbelt, MD 20771, USA

# ABSTRACT

Studies on Voyager 1 using the CRS instrument have shown the presence of sub-MeV electrons in the interstellar medium beyond the heliopause. We believe that these electrons are the very low energy "tail" of the distribution of galactic GeV cosmic ray electrons produced in the galaxy. If so this observation places constraints on the origin and possible source distribution of these electrons in the galaxy. The intensities of these electrons as well as MeV protons and other higher energy electrons and nuclei have been followed outward from the Earth to beyond the heliopause during the 40 years of the Voyager mission. Among the other new features found in this study of the radial dependence of the electron intensity in the heliosphere are: (1) The heliosheath is a source of sub-MeV electrons as well as the already known anomalous cosmic rays of MeV and above, none of which appear to escape from the heliosphere because of an almost impenetrable heliopause at these lower energies; (2) Solar modulation effects are observed for these MeV electrons throughout the heliosphere. These modulation effects are particularly strong for electrons in the heliosheath and comprise over 90% of the observed intensity change of these electrons of 10-60 MeV between the Earth and the heliopause. Even for nuclei of ~1 GV in rigidity, over 30% of the total intensity difference between the Earth and the LIM occurs in the heliosheath; (3) The 2 MeV protons studied here for the first time beyond the heliopause are also part of the low energy tail of the spectrum of galactic cosmic ray protons, similar to the tail noted above for sub MeV galactic cosmic ray electrons.

### Introduction

This study is mostly in uncharted territory and parallels a recent study of much higher energy H and He nuclei (~1 GV rigidity) using the CRS telescopes (Stone, et al., 2013) on Voyager 1 and 2 (V1-V2) (Webber, et al., 2017). This earlier study of nuclei followed the intensities of ~250 MeV/nuc H and He nuclei throughout the heliosphere inside the heliospheric termination shock (HTS) and beyond in the heliosheath and then into the LIS region beyond the heliopause (HP) where their intensities have remained essentially constant to within $\pm$ 1% over an additional 20 AU of travel beyond the heliopause. For protons ~250 MeV (0.72 GV) the intensity increased by ~8 times from the Earth in 1977 to LIS space in 2012. The well-known 11 year solar modulation could be followed for three 11 year solar cycles out to the HTS, with maxima in intensity in 1977, 1987 and 1998. Beyond the HTS a considerable modulation of magnitude at least 1/3 of the total modulation observed at the Earth, was present in the heliosheath although the solar cycle variations were less obvious.

For the electrons used in this paper we make use of two energy channels, including a reference energy ~4 MeV made using the HET telescope in the B stop electron mode (see Cummings, et al., 2016). A low background electron rate has also been found serendipitously in the HET-A stop telescope rates, e.g., A1·A2^C1. This new study has been enabled by further examination of unidentified clumps of particles in the low pulse height corner of the matrices of events A1 vs. A2 in the A1·A2^C1 telescope similar to what was seen earlier on the B1·B2^C4 matrices; and also by using GEANT 4 calculations of the response of these counters to electrons which helped to identify these clumps. The A1·A2^C1 telescope was originally believed to be non-responsive to electrons at these low energies, but the GEANT 4 calculations identify a response, sharply defined in energy between ~0.2-0.7 MeV. The double dE/dx measurement provides a low background. This response could only be possible if the thresholds in these counters on V1 were lower than originally believed. Studies of the absolute intensities of these sub MeV electrons and a comparison of V1 and V2 intensities are ongoing and will be reported in subsequent publications.

The two electron energy bands, from a few tenths of an MeV up to a few MeV, have been continuously monitored since the launch of V1 in 1977. This time period includes nearly 3 solar 11 year activity cycles inside the HTS and an additional 8 years, corresponding to ~27 AU of outward travel in the heliosheath, where a reduction of intensity by a factor of up to 100 between the LIS electron intensity and the intensity observed at the HTS boundary, due to solar modulation effects, has recently been determined for electrons between 10-60 MeV (Webber, et al., 2018). The LIS intensities of electrons beyond the HP at both of these energies, above and below 1 MeV, have remained constant to within a few percent for over 5 years (~20 AU of outward travel for V1) in this region beyond the heliopause and are believed to be the low energy part of the galactic cosmic ray electron spectrum.

This relatively constant LIS component, for both electrons and protons, is in contrast to the large variabilities observed at all energies for both electrons and protons within the heliosphere.

This constancy of intensity is observed for all components, electrons and protons, so we will use this LIS intensity measured for each component as a reference for the studies reported here.

## **The Radial Profiles**

### **a) Low Energy Electron Profiles**

Figure 1 shows the 26 day average rates of the two lowest energy electron channels, the HET A1-A2 stop channel at ~0.5 $\pm$ 0.3 MeV and the HET B1-B2-C4=1 and 2 channels with a peak response between 3 and 5 MeV, from the time of launch (1977) to the present time (2018). These responses are determined by extensive GEANT-4 calculations. These telescopes are described in Stone, et al., 1977, and Cummings, et al., 2016. The rates are normalized to those measured for A1·A2 stop in LIS space.

One can define from Figure 1 four temporal or spatial regions that apply to all the components and energies discussed here. The 1$^{st}$ region is from launch in 1977 to 1983.0 when V1 was at about 20 AU and also near its maximum latitude of ~30°. This region nearest the Sun is punctuated by many abrupt sub MeV electron increases related to interplanetary propagating events of solar origin. These Voyager measurements are closely correlated with corresponding observations of 0.5 MeV electrons at Helios inside the Earth's orbit during the same 1977-78 time period.  The individual events in this time period are separated temporally as is seen in the daily average rates presented later in Figure 5.

The second region (time period) is from 1983.0 to 2005.0 at which time V1 crossed the HTS and entered the heliosheath at ~95.0 AU at ~30° N latitude.  The effects of the solar 11 year modulation cycle are rather weakly seen in both electron profiles in this region.  The temporal variations of these electrons are similar, but much weaker than what is observed for higher energy protons (Webber, et al., 2017) and shown later in Figure 2.  Two large solar induced interplanetary events are seen in the 0.5 MeV rates in 1989 and 1991 when V1 was at between 40 and 50 AU, the furthest distance at which shock accelerated electrons are observed.

During the time V1 is in the heliosheath region (Region 3) from 2005.0 to 2012.65, between 95 and 121.7 AU, there is a factor of 3-4 increase in the sub-MeV electron intensities which, during most of this time period, have even higher intensities than those observed beyond the HP, in the local interstellar medium.  In contrast the 3-5 MeV electrons increase beyond the HTS in the heliosheath by a factor ~30 reaching a maximum at the LIS intensity.  This increase is believed to be a result of modulation effects in the heliosheath and the absence of the heliosheath accelerated electrons which are present only in the lower energy channel.

Beyond the HP in the LIM the electron intensities have remained constant to within $\pm 5\%$ at both energies for over 5 years (~20 AU or outward travel for V1).  Notice that the intensities in the LIS are higher than those observed at the Earth in 1977 by a factor ~4 at 0.5 MeV and a factor ~20 at 4 MeV.  These differences are due in part to solar modulation effects.

Figure 2 shows the 0.5 and 3-5 MeV electron rates in Figure 1 on an expanded time scale covering the time period from 2002 to 2015 when V1 was passing through the heliosheath.  Note that for the higher energy electrons the rate increases continuously from the HTS to the HP in the heliosheath by a factor ~30.  We believe that this is due to solar modulation effects.

For the sub MeV electrons there are two "precursor" events (more weakly seen at higher energies) followed by an increase at the HTS at 2005.0.  The intensity time profile of these electrons in the heliosheath is almost an exact match to the profiles of higher energy anomalous particles accelerated in the heliosheath, suggesting a common origin.

The total increase of the sub MeV and 3-5 MeV electrons between just inside the HTS and just outside the HP is greatly different, however, ranging from a factor ~2.5 for sub MeV electrons to a factor ~30 as noted earlier for the 3-5 MeV electrons, a difference of a factor ~12. Outside the HP the difference in the two intensities is consistent with a galactic spectrum ~$E^{-1.35}$ between 0.5 and 4 or a factor of about 16. So inside the HTS this intensity difference between the two energies is 16 x 12 = ~192. This factor would be indicate a spectrum~$E^{-2.50}$ inside the HTS between these two low energies. Thus, if the electron spectrum inside the HTS has, as its origin, the sub MeV electrons measured in the heliosheath then, based on this sub MeV spectrum, they would not be observed in the 3-5 MeV intensity profile observed inside the HTS, as is the case.

### b) Electron vs. Nuclei Profiles

In Figure 3 we show the higher energy (250 MeV = 720 MV rigidity) proton intensity time profile from the Webber, et al., 2017, solar modulation study. This proton intensity is also normalized to the 0.5 MeV A1-A2 stop LIS electron rate beyond the HP.

The 11 year solar modulation is observed in region 1 inside 20 AU where the Voyager observations of 250 MeV protons blend in well with spacecraft observations near the Earth and also with neutron monitor observations (Mewaldt, et al., 2010, Lave, et al., 2013 and Usoskin, et al, 2011) at the same time. The intensity at the Earth in 1977 is a factor ~8 below the LIS value at this energy.

This solar modulation of these higher energy protons can be well described with a single modulation parameter $\phi$ in MV corresponding to an energy loss resulting from a potential difference, $\phi$ in MV, between the observation point and the LIS spectrum (e.g., Gleeson and Axford, 1998). This simple description arises from the fact that the overall spherically symmetric solar modulation in the heliosphere appears to follow the description provided by Louvilles Theorem relating to the constancy of the particle density and momentum in phase space. Of course there are significant deviations from this simple picture due to structural features in the heliosphere such as the tilt of the heliospheric current sheet, and also for the solar polarity changes which induce a 22 year cycle in the solar modulation process, but these other processes do not appear to dominate at these higher energies and above, where, in fact, the same value of the modulation potential, $\phi$, obtained from the Voyager studies also gives a good description of the historical neutron monitor observations of cosmic ray modulation effects at the Earth that have been carried out over the last 70 years.

In region 2, between about 20-95 AU, solar 11 year modulation effects are observed at 250 MeV in synch in both time and magnitude with those observed at the Earth, with a time delay corresponding to outward moving structural features with a typical average speed ~600 km·s$^{-1}$.

Moving outward into region 3, we find that the intensity of these 250 MeV protons increases by a factor ~4 in the heliosheath. The large modulation beyond the HTS was previously unrecognized and is a significant fraction (~30%) of the total modulation observed at the Earth. The individual solar 11 year modulation effects are not as evident in the heliosheath.

And next, in Figure 4, we show the 1.8-2.6 MeV proton rate, again normalized to the LIS rate that is measured for 0.5 MeV electrons. This is the lowest energy part of the interstellar proton spectrum that can be measured accurately on Voyager in a 2-D, dE/dx, telescope mode. It represents a limit to the characteristics of the lowest energy proton propagation in the galaxy from the nearest sources to the Sun since it corresponds to a range of only ~35$\mu$ of equivalent Si traversed during the galactic propagation.

In region 1, between launch and 1983 and inside ~20 AU, the Earth and the inner heliosphere are bathed in an almost continuous flux of these 2 MeV protons exceeding the possible background galactic cosmic rays at this energy and location by a factor ~100 at times and comparable to the proton intensities observed further out in the heliosheath. This continuously high intensity in the inner heliosphere results from the 1.8-2.6 MeV protons from many individual events overlapping because of azimuthal broadening of the profiles.

In region 2, between ~20 AU and the HTS at 95 AU, the protons reach a low intensity level which is ~0.1 of that in the LIS. This corresponds to a solar activity minimum and also a solar modulation minimum. This intensity, which is the lowest observed by Voyager 1, and which is a factor ~12 times less than the LIS intensity, could be a representation of the level of solar modulation at this low energy of ~2 MeV.

In the heliosheath in region 3, the intensity of these low energy protons increases by a factor ~1000 beyond the HTS from that observed in region 2. These low energy protons are accelerated in the heliosheath; then after 8 years and 28 AU of outward propagation, their intensity decreases suddenly by a factor ~500 to the LIS value in only 1-26 day interval corresponding to a distance of ~0.1 AU as V1 crosses the heliopause. This intensity change over such a small radial interval requires a remarkably effective particle boundary.

Here we summarize some of the most general features of all three of the radial intensity profiles that have been presented. We believe that one of the most prominent features in all cases is the sharpness and effectiveness of the heliospheric boundary, the heliopause. For low energy protons the reduction of intensity at this boundary is a factor ~500, taking place in only 1-26 day interval corresponding to less than 0.1 AU in distance as noted above.

The effects of this boundary on low energy electrons are even more astounding. A radial intensity gradient 130%/AU just inside the HP for 4 MeV electrons changes to ~0.1%/AU gradient in the LIM beyond the heliopause also over a 26 day time interval of ~0.1 AU in radius. Essentially whatever intensities exist inside the HP, apparently stay in the heliosphere as a result of an almost impenetrable heliopause at these lower energies. As far as energetic particles up to tens of MeV/nuc go, the interstellar medium at this location has little recognition of the nearby heliosphere.

Other features of the heliosphere newly recognized from this study include: 1) The heliosheath is a very interesting and important region both in terms of accelerating protons and heavier nuclei, as is well known, but also sub MeV electrons. These electrons have an intensity profile very similar to that of anomalous cosmic rays of tens of MeV/nuc suggesting a possible common origin.

In addition large solar modulation effects of electrons are observed in the heliosheath. The intensity modulation effects in this region range from a factor ~4 for ~1 GV protons to a factor ~100 for 15 MV electrons and then decreasing to a factor ~30 for 4 MV electrons and much less for sub MeV electrons. The effects of solar modulation studied now for over 70 years at or near the Earth using neutron monitors, balloon and spacecraft borne instruments at higher energies, and so important for geophysical studies (e.g., Beer, McCracken and von Steiger, 2013), are still observed for the higher energy particles beyond the HTS in the heliosheath.

The new observations reported here have extended our measurements of the interstellar electron spectrum by a factor ~10 lower in energy (e.g., down to 0.5 MeV) as compared with even the initial Voyager measurements beyond the HP of Stone, et al., in 2013. For protons the minimum energy is now ~1.8 MeV, a factor ~2 times lower than the initial measurement for these particles. The intensities obtained at these lowest energies beyond the HP appear to be consistent with a smooth extension of the galactic cosmic ray measurements at higher energies as would be expected from galactic propagation models. The range of the lowest energy electrons and protons is only ~35 $\mu$ of Si. This places severe constraints on the distribution of nearby sources of these particles but also illustrates that, even at these lower energies, the observed spectra are still a compendium of many sources with an individuality obscured by diffusion in the galactic magnetic fields.

**Daily Average Intensity Time Profiles of ~0.5 MeV Electrons**

Daily average data is also available for the 0.5 MeV electrons and is very valuable for the study of shorter term temporal variations. In Figure 5 we show the daily average rate for the time period from launch in 1977 to the end of 1981. This includes the Jupiter and Saturn encounters in 1979 and 1980 respectively, as well as several outward propagating events of solar origin. These events are well separated in time and space and can be directly compared with similar 0.5 MeV electron data on Helios and also on V2 to understand the aspects of both radial and azimuthal diffusion.

Note that the events in 1981 have much broader intensity-time profile extending to ~26 days.

**Summary and Conclusions**

The most important features of the data have been summarized in the above text. Here we consider again the 0.5 MeV electrons. First and foremost is the intensity of these electrons beyond the heliopause. Preliminary intensity estimates in P/m$^2$·sr·s/MeV of the 0.5 MeV intensity using the GEANT-4 program are consistent with an extension to lower energies of the $E^{-1.35}$ spectrum measured at Voyager 1 (Cummings, et al., 2016) for these interstellar electrons between 3-60 MeV.

A significant excess of these sub MeV electrons is also observed in the heliosheath. This excess is not observed for electrons >3 MeV, suggesting that the spectra of these lower energy electrons is steep.

These 0.5 MeV electrons are also seen as a "background" throughout interplanetary space inside the HTS, interspersed with solar related shocks (CME) propagating into the outer most regions inside the HTS with possible enhancements as they approach the HTS. This "background" is currently being evaluated and we believe now that a large fraction of the 0.5 MeV electron rate is due to true electrons.

**<u>Acknowledgements:</u>** The authors are grateful to the Voyager team that designed and built the CRS experiment with the hope that one day it would measure the galactic spectra of nuclei and electrons. This includes the present team with Ed Stone as PI, Alan Cummings, Nand Lal and Bryant Heikkila, and to others who are no longer members of the team, F.B. McDonald and R.E. Vogt. Their prescience will not be forgotten. This work has been supported throughout the more than 40 years since the launch of the Voyagers by the JPL.

## FIGURE CAPTIONS

**Figure 1:**  26 day average rates of 0.5 MeV and 3-5 MeV electrons on V1 from launch in 1977 to 2018.  The rates are normalized to the LIS rate of 0.22 c/s for 0.5 MeV electrons.  The radial distance from the Sun is from 1 to 140 AU with the HP crossing at 2012.6 at 121.7 AU and the HTS crossing at 2004.9 at 95 AU.

**Figure 2:**  The 0.5 and 3-5 MeV electron rates shown in Figure 1 on an expanded time scale

**Figure 3:**  Same as Figure 1 but for the LIS 250 MeV proton intensity normalized to the LIS 0.5 MeV electron rate.

**Figure 4:**  Same as Figure 1 but for the LIS 2.0 MeV proton rate normalized to the LIS 0.5 MeV electron rate.

**Figure 5:**  Daily average 0.5 MeV electron rate at Voyager 1 from launch to the end of 1981.  The planetary encounters with Jupiter and Saturn are indicated with a J or S.  The other numbered increases are solar originating shock structures moving outward from the Sun.

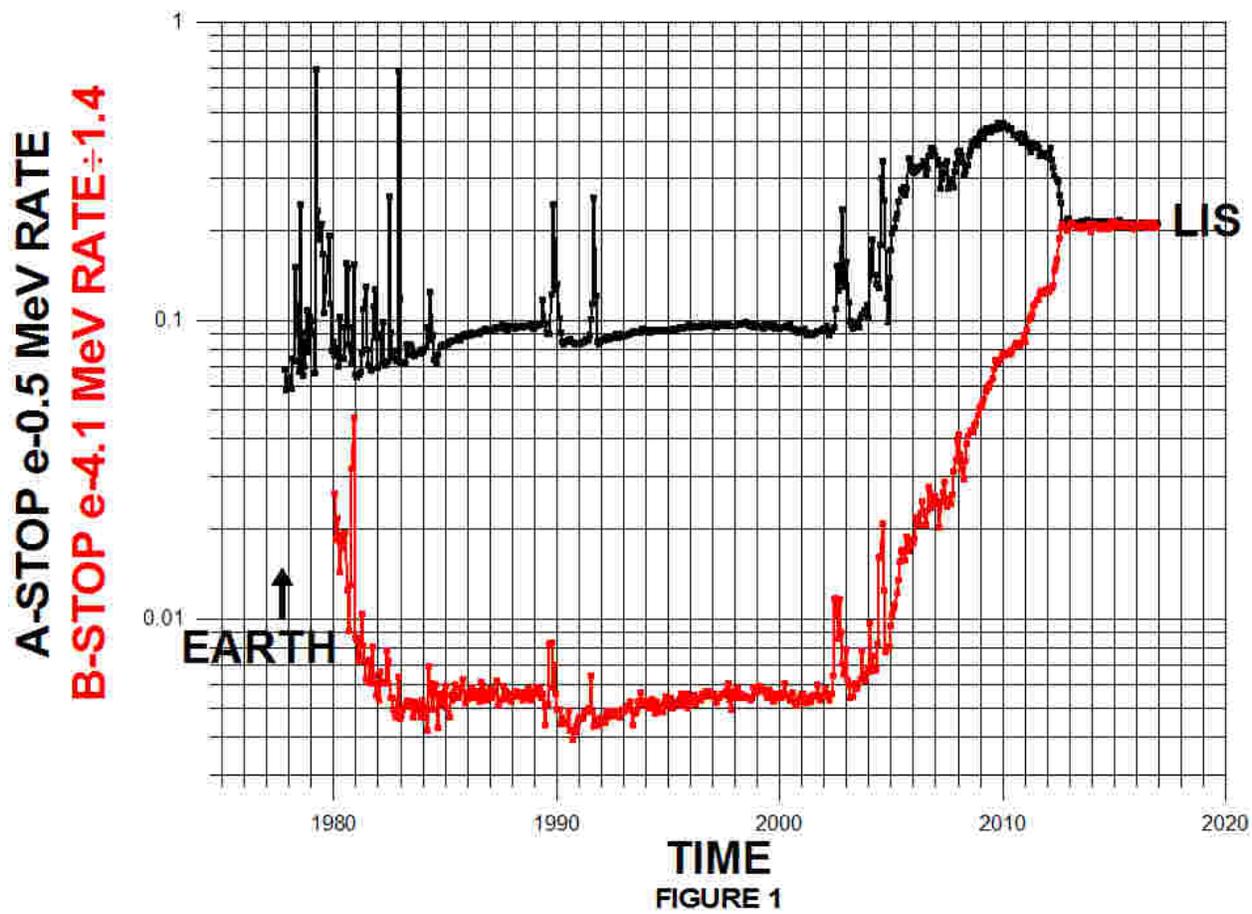

**FIGURE 1**

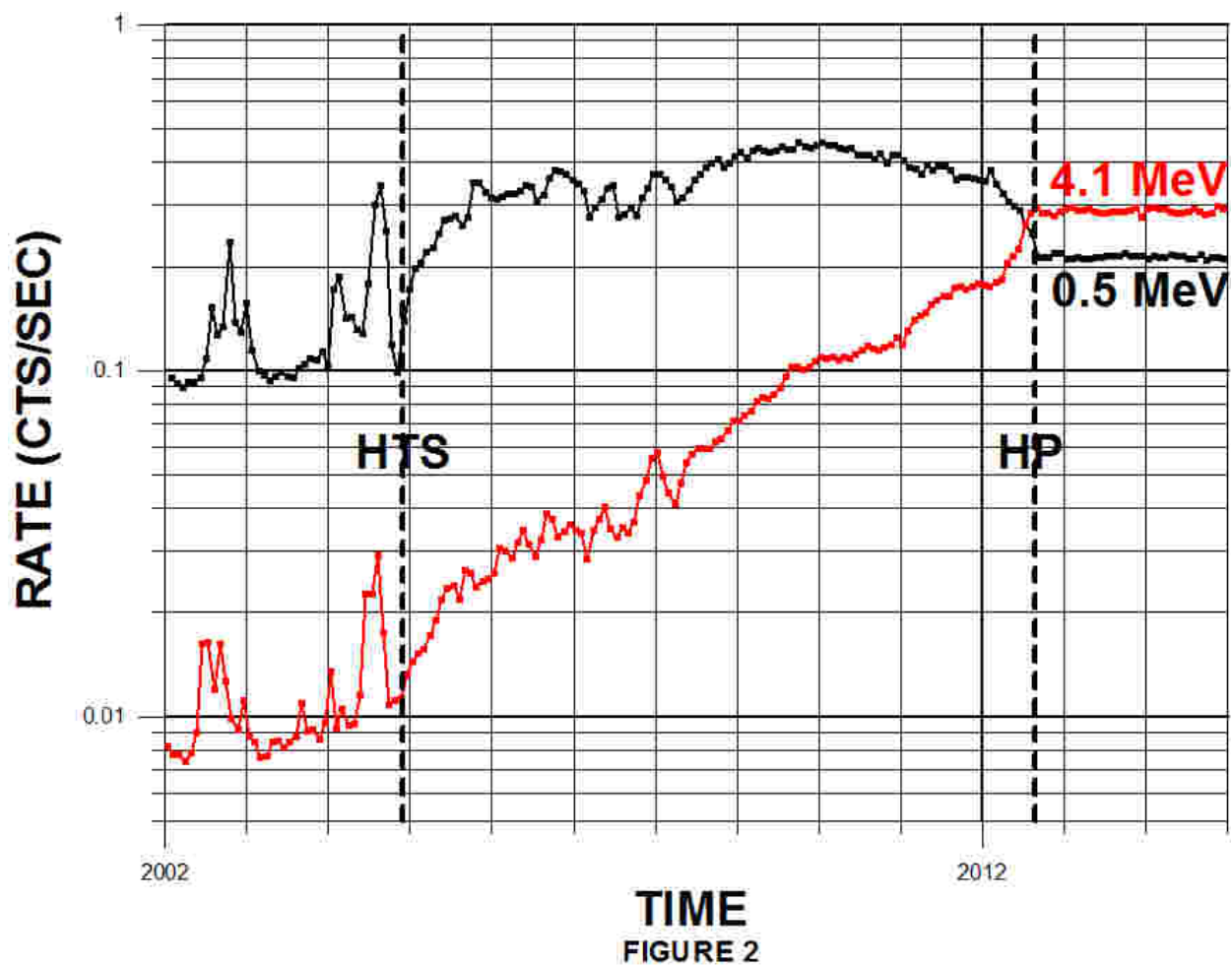

**FIGURE 2**

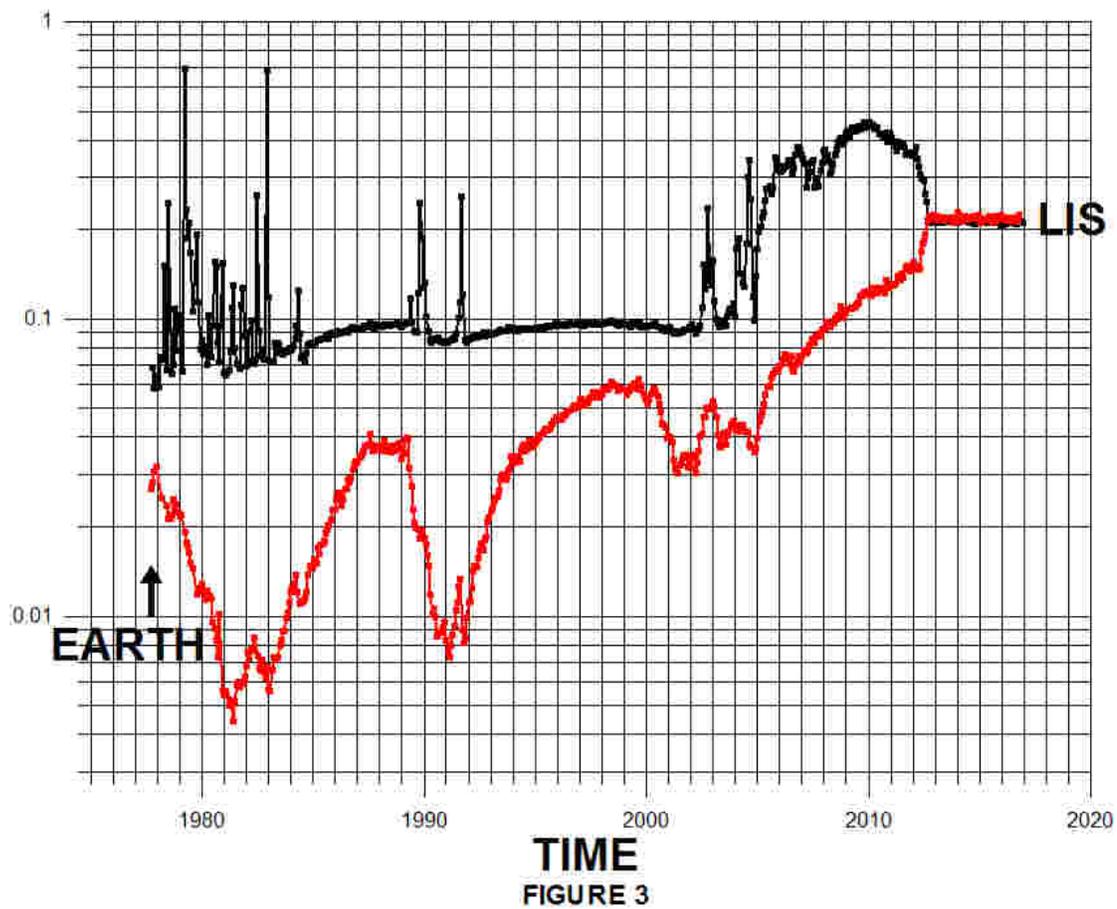

**FIGURE 3**

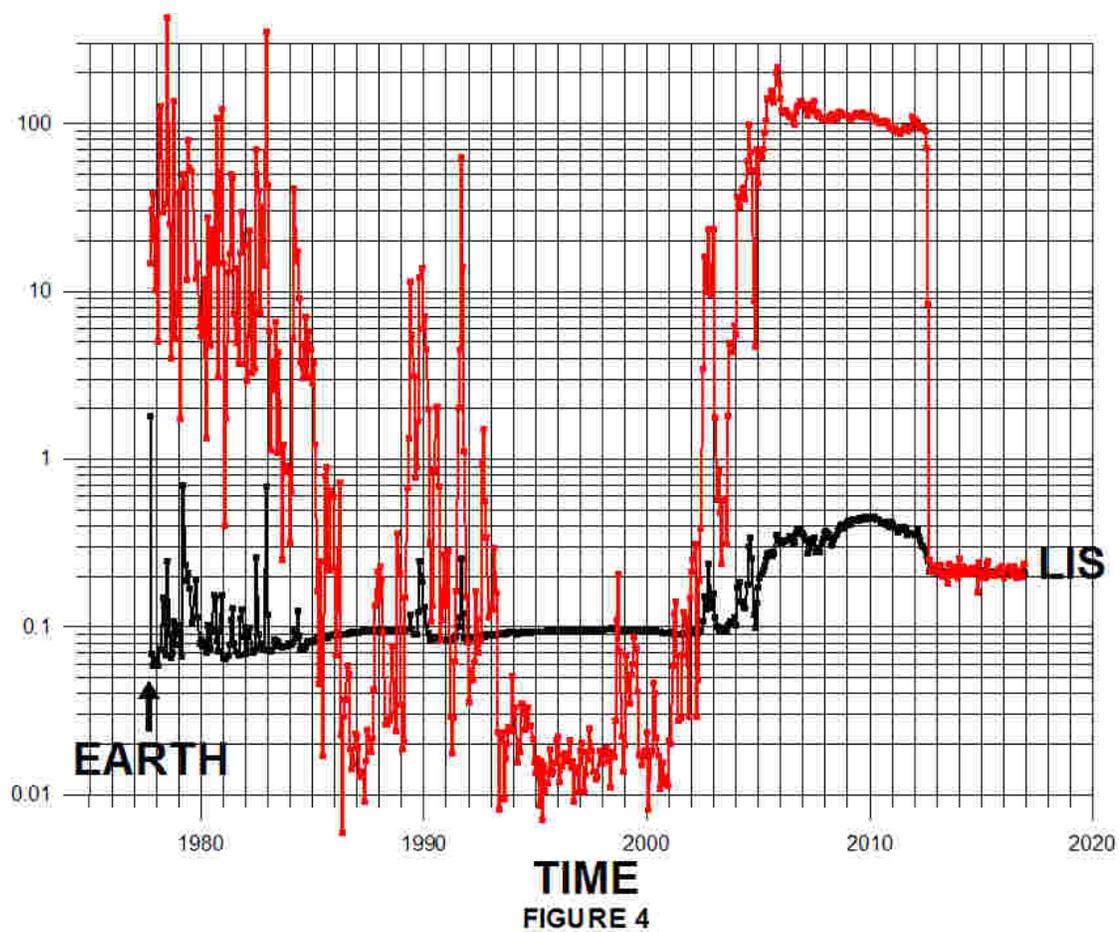

**FIGURE 4**

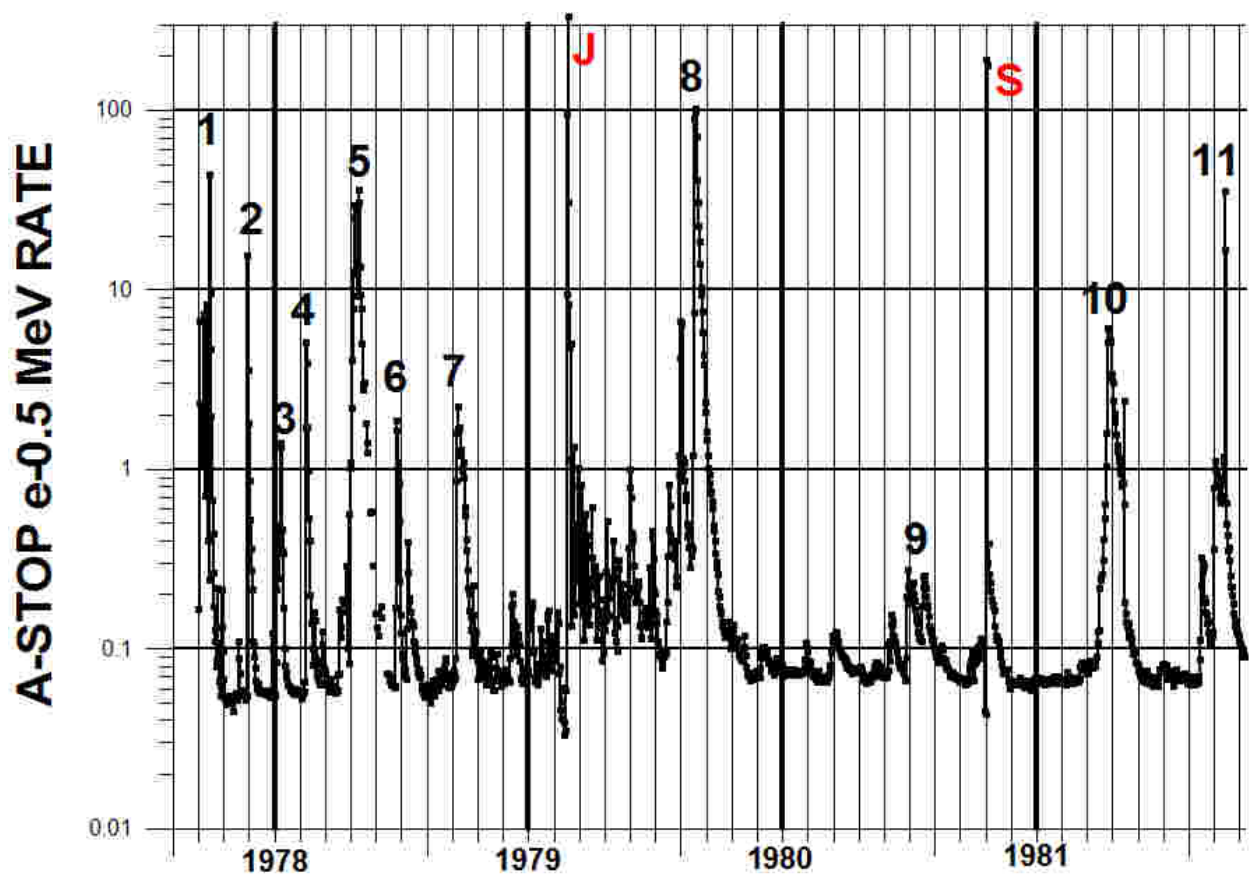

FIGURE 5